\documentstyle[11pt]{article}\textheight 230mm\textwidth 150mm
            \newcommand{\be}{\begin{eqnarray}}
            \newcommand{\ee}{\end{eqnarray}}
            \newcommand{\eel}[1]{\label{#1}\end{eqnarray}}
\newcommand{\e}[1]{\label{e:#1}\end{eqnarray}}
     \newcommand{\eg}{{\em e.g.\ }}
            \newcommand{\ie}{{\em i.e.\ }}
            \newcommand{\ga}{{\gamma}}
 
            \newcommand{\la}{{\lambda}}
            
            \newcommand{\del}{{\delta}}

           \newcommand{\ra}{{\rightarrow}}
 \newcommand{\lea}{{\leftarrow}}

            \newcommand{\pet}{{\cal P}}

\newcommand{\ca}{{\cal C}}

            \newcommand{\beq}{\begin{quote}}
            \newcommand{\eq}{\end{quote}}
            \newcommand{\Om}{\Omega}
   
            \newcommand{\al}{\alpha}
            \newcommand{\ben}{\begin{enumerate}}
            \newcommand{\een}{\end{enumerate}}
            \newcommand{\bit}{\begin{itemize}}
            \newcommand{\ei}{\end{itemize}}
    	\newcommand{\nn}{\nonumber}
            \newcommand{\r}[1]{(\ref{e:#1})}
            \newcommand{\edfl}[1]{\label{#1}\end{df}}
\newcommand{\vb}{{\cal h}}
\newcommand{\hb}{{\cal i}}

\newcommand{\ve}{{\varepsilon}}

\newcommand{\bett}{{\bf 1}}
	
	\newcommand{\dif}{\partial}

  \def\half{{1 \over 2}}

\def\JMP{{\sl J.\ Math.\ Phys.}}
\begin{document}
\begin{titlepage}
\noindent
September 2000\\

\vspace*{5 mm}
\vspace*{35mm}
\begin{center}{\LARGE\bf Open group  transformations\footnote{Talk
presented at the conference
``Quantization, Gauge Theory, and Strings" dedicated to the memory of
Professor Efim Fradkin,
Moscow, June 5-10, 2000}}
\end{center} \vspace*{3 mm} \begin{center} \vspace*{3 mm}

\begin{center}Robert
Marnelius\footnote{E-mail: tferm@fy.chalmers.se.}\\
\vspace*{7 mm} {\sl Institute
of Theoretical Physics\\ Chalmers University of
Technology\\ G\"{o}teborg
University\\ S-412 96  G\"{o}teborg, Sweden}\end{center}
\vspace*{25 mm}
\begin{abstract}
Recent results on finite open group transformations are reviewed.
\end{abstract}\end{center}\end{titlepage}

\setcounter{page}{1}
\noindent
{\bf Introduction}\\ \\
It is a pleasure for me  to be  at this conference dedicated
to the memory of Efim
Fradkin. I met Fradkin for the first time at Alushta, Crimea
in April 1976 where he gave a talk on
general constraints and their quantization. This memory is
particularly pleasant to me now since the
content of my  talk is closely related to the highlights of
Fradkin's general constraint theory. My  talk is a review of 
 some recent work I have done together
with Igor Batalin, one of Fradkin's
most frequent collaborators on these matters who unfortunately
was unable to deliver a talk at this
time due to family reasons. I
will talk about finite open  group transformations and then mainly as gauge transformations within the Batalin-Fradkin-Vilkovisky (BFV) framework.
It is based on four papers
\cite{BM1}-\cite{BM4} with emphasis  on the second one
which has the same title as this talk.

 Let me first define
what I mean by open
groups.
Consider a Hamiltonian formulation on a symplectic manifold
$\Gamma$ with coordinates $z^A$.
Let $\theta_{\al}(z)$ be functions on $\Gamma$ with Grassmann parities,
$\ve(\theta_{\al})\equiv\ve_{\al}=0,1$, satisfying the Poisson algebra
\be
&&\{\theta_{\al}(z),
\theta_{\beta}(z)\}=U_{\al\beta}^{\;\;\ga}(z)\theta_{\ga}(z),
\e{1}
where $U_{\al\beta}^{\;\;\ga}(z)$ are restricted by the Jacobi identities. For
$U_{\al\beta}^{\;\;\ga}$ constants
$\theta_{\al}(z)$ are generators of a Lie group while for
$U_{\al\beta}^{\;\;\ga}(z)$  functions
$\theta_{\al}(z)$ are generators of open groups
(sometimes called groupoids). The transformations
generated by $\theta_{\al}(z)$ are symmetry transformations in general only if
$\theta_{\al}(z)=0$ at the level
of the equations of motion. Open groups as symmetry groups
are therefore in general gauge groups. Important
open gauge groups appear in gravity, supergravity and p-branes.
In this talk I will show that it is
possible to integrate \r{1}. I will explicitly construct an
exponential representation of the
group elements. This will be  done directly at the quantum level.
On our way we will find some
remarkable mathematics.\\ \\
{\bf The quantum Lie algebra - the BFV-BRST charge}\\ \\
The first question we have to answer is how   \r{1} should
be quantized in order to
preserve the Lie algebra. The answer is given by the
general BRST formulation given by Batalin,
Fradkin and Vilkovisky -- the BFV-formalism \cite{BFV}.
They require us  first to ghost extend the
manifold by generalized Faddeev-Popov ghosts $\ca^{\al}$
and their canonical momenta  $\pet_{\al}$
with Grassmann parities $\ve(\ca^{\al})=\ve_{\al}+1$,
$\ve_{\al}\equiv\ve(\theta_{\al})$. Then they require
us to construct the odd  BFV-BRST charge
\be
&&\Om=\ca^{\al}\theta_{\al}(z)+\half
\ca^{\beta}\ca^{\al}U_{\al\beta}^{\;\;\ga}\pet_{\ga}
(-1)^{\ve_{\ga}+\ve_{\beta}}+\ldots,
\e{2}
 where the dots indicates terms of higher powers
in the ghost momenta. These terms are determined by
the condition
\be
&&\{\Om, \Om\}=0,
\e{3}
which is equivalent to \r{1} together with all
its Jacobi identities. That this always is possible
is, I think,  Fradkin's most important contribution to the
constraint theory. The maximal power of
$\pet_{\al}$ in
$\Om$ defines the rank of the group which is a way
to roughly classify open groups. For rank$\geq2$
we have open groups. However, even rank one contains
nontrivial open groups, the quasi groups
\cite{IB}. Batalin and Fradkin have shown that for
finite dimensional manifolds with globally
defined canonical coordinates it is always possible
to quantize such that $\hat{\Om}^2=0$ where
$\hat{\Om}$ is the corresponding operator defined up to
$\hbar$-corrections \cite{BF}. (For infinite degrees of
freedom we may have anomalies.) $\hat{\Om}$
in
 $\ca\pet$-ordered form is (``hat" is suppressed in the following)
\be
&&\Om=\sum_{i=0}^N\Om'_i,\;\;\;\Om'_0\equiv\ca^{\al}\theta'_{\al},\nn\\
&&\Om'_i\equiv \ca^{{\al}_{i+1}}\cdots\ca^{{\al}_1}{\Om'}_{{\al}_1\cdots
{\al}_{i+1}}^{{\beta}_i\cdots {\beta}_1}\pet_{{\beta}_1}\cdots\pet_{{\beta}_i},
\e{4}
where $\theta'_{\al}=\theta_{\al}+\hbar$-corrections.
$\Om^2=0$ requires then
$[\theta'_{\al},
\theta'_{\beta}]=i\hbar{U'}_{{\al}{\beta}}^{\;\;\;{\ga}}\theta'_{\ga}$,
where
${U'}_{{\al}{\beta}}^{\;\;\;{\ga}}=2(-1)^{\ve_{\beta}+\ve_{\ga}}
{\Om'}_{{\al}{\beta}}^{\;\;\;{\ga}}=
{U}_{{\al}{\beta}}^{\;\;\;{\ga}}+\hbar$-corrections.
Thus, the operator ${\Om}$ indeed represents
 the  quantum Lie algebra if $\Om^2=0$.\\ \\
{\bf A scenario for possible forms of group elements}\\ \\
I am now going to integrate the Lie algebra inherent
in the BFV-BRST charge operator $\Om$.
However, let me  first give a scenario for finite
open group transformations which is exactly true
for Lie group theories before I embark on the actual
derivation. Imagine we have a finite group
element
$U(\phi)$ where
$\phi^{\al}$ are group parameters,
$\ve(\phi^{\al})=\ve_{\al}\equiv\ve(\theta_{\al})$.
It is natural to expect the existence of an
exponential representation
\be
&&U(\phi)=\exp{\{{i\over\hbar}F(\phi)\}},
\e{5}
where $F(0)=0$ by a choice of parametrization
($U(0)=\bett$), and where
$[\Om, F(\phi)]=0$ which is  required by
applications to gauge groups. We also assume $[G,
F(\phi)]=0$ where $G$ is the ghost charge
($[G, \Om]=i\hbar\Om$) so that group transformations do
not change ghost numbers.
Since  $U(\phi)$ performs gauge transformations in gauge theories we have
\be
&&\vb ph|U(\phi)|ph\hb=\vb ph|ph\hb,
\e{6}
 when $\Om|ph\hb=0$.
This requires
\be
&&F(\phi)={1\over i\hbar}[\Om, \rho(\phi)].
\e{7}
Furthermore, since $\tilde{\theta}_{\al}\equiv(i\hbar)^{-1}
[\Om, \pet_{\al}](-1)^{\ve_{\al}}$
represent group generators within a BRST framework for
Lie group theories \cite{RM} we expect
$\rho(\phi)=\pet_{\al}\phi^{\al}$. The above scenario
is exactly true for theories with rank zero
and one. However, for open groups of  rank two and higher
$\tilde{\theta}_{\al}$ satisfy a closed algebra only together
 with  $\pet_{\al}$. This means that
the final picture in this case must be somewhat different,
which indeed is the case as we shall see.
Let us now turn to the actual derivation.\\ \\
{\bf Actual derivation}\\ \\
The starting point in our derivation is the Lie equations
\be
&&\vb A(\phi)|\stackrel{\lea}{D}_{\al}\equiv\vb
A(\phi)|\left(\stackrel{\lea}{\dif_{\al}}-
(i\hbar)^{-1} Y_{\al}(\phi)\right)=0,\nn\\
&&A(\phi)\stackrel{\lea}{\nabla}_a\equiv A(\phi)
\stackrel{\lea}{\dif_a}-(i\hbar)^{-1}
[A(\phi), Y_a(\phi)]=0,
\e{8}
where $Y_{\al}(\phi)$ is a connection operator.
The integrability
conditions for $Y_{\al}(\phi)$ are
\be
&&Y_{\al}\stackrel{\lea}{\dif_{\beta}}-Y_{\beta}
\stackrel{\lea}{\dif_{\al}}(-1)^{\ve_{\al}\ve_{\beta}}=
(i\hbar)^{-1}[Y_{\al},
Y_{\beta}].
\e{9}
The above scenario then suggests that
\be
&&Y_{\al}(\phi)=(i\hbar)^{-1}[\Om, \Om_{\al}(\phi)],
\e{10}
where $\Om_{\al}(0)=\pet_{\al}$.
The integrability conditions \r{9} for $Y_{\al}$ imply then
\be
&&[\Om, \Om_{\al}\stackrel{\lea}{\dif_{\beta}}-\Om_{\beta}
\stackrel{\lea}{\dif_{\al}}(-1)^{\ve_{\al}\ve_{\beta}}-
(i\hbar)^{-2}(\Om_{\al}, \Om_{\beta})_{\Om}]=0,
\e{11}
where I have introduced \\ \\
{\bf The quantum antibracket}\\ \\
The quantum antibracket is defined by \cite{BM5,BM6}
\be
&&(f, g)_Q\equiv \half \left([f, [Q, g]]-
[g, [Q, f]](-1)^{(\ve_f+1)(\ve_g+1)}\right),
\e{12}
where $Q$ is any odd operator. The quantum
antibracket is  a new general algebraic tool in terms of
which one \eg may  give an operator version of the
BV-quantization of general gauge theories
\cite{BM5,BM6}. It generalizes the classical antibracket
to operators. In fact, it reduces to the
classical antibracket when $f$ and
$g$ are functions and when
$Q$ is a second order nilpotent differential operator.
The quantum antibracket satisfies the
algebraic properties of the classical antibracket except
that we have a generalized Leibniz' rule:
\be
&&(fg, h)_Q-f(g, h)_Q-(f, h)_Qg(-1)^{\ve_g(\ve_h+1)}=\nn\\
&&\half\left([f, h][g,
Q](-1)^{\ve_h(\ve_g+1)}+[f,Q][g,h](-1)^{\ve_g}\right).
\e{13}
The ordinary Leibniz' rule is only satisfied for commuting operators.
The quantum antibracket also
satisfies generalized Jacobi identities. For $Q^2=0$ we have
\be
&&(f,(g, h)_Q)_Q(-1)^{(\ve_f+1)(\ve_h+1)}+cycle(f,g,h)=\nn\\
&&-\half[(f, g,
h)_Q(-1)^{(\ve_f+1)(\ve_h+1)}, Q],
\e{14}
where $(f, g, h)_Q$ is an example of higher
quantum antibrackets defined by \cite{BM6}
\be
&&(f_{a_1},\ldots, f_{a_n})_Q\equiv -
\left.Q(\la)\stackrel{\lea}{\dif}_{a_1}\stackrel{\lea}{\dif}_{a_2}\cdots
\stackrel{\lea}{\dif}_{a_n}(-1)^{E_n}\right|_{\la=0},
\e{15}
where the derivatives, ${\dif}_{a}$, are with respect
to the parameters, $\la^a$, and where
\be
&&Q(\la)\equiv e^{-A}Qe^A,\quad A\equiv f_a\la^a, \quad E_n\equiv
\sum_{k=0}^{\left[{n-1\over 2}\right]}\ve_{a_{\rm 2k+1}},
\e{16}
where in turn $\ve_a\equiv\ve(f_a)=\ve(\la^a)$.
\newpage
\noindent
{\bf Back to the integrability conditions \r{11}}\\ \\
The integrability conditions \r{11} for $\Om_{\al}$
 may be cast into the form
\be
&&\Om_{\al}\stackrel{\lea}{\dif_{\beta}}-\Om_{\beta}
\stackrel{\lea}{\dif_{\al}}(-1)^{\ve_{\al}\ve_{\beta}}-
(i\hbar)^{-2}(\Om_{\al}, \Om_{\beta})_{\Om}+\half
(i\hbar)^{-1}[\Om_{{\al}{\beta}}, \Om]=0.
\e{17}
That there must be a BRST exact solution of \r{11} is
due to the fact that $\Om_{\al}$ has negative
ghost number.  From these integrability conditions  we may then derive
integrability conditions for $\Om_{{\al}{\beta}}$ given by
\be
&&\dif_{\al}\Om_{\beta\ga}(-1)^{\ve_{\al}\ve_{\ga}}+\half(i\hbar)^{-2}(\Om_{
\al},
\Om_{\beta\ga})_{\Om}(-1)^{\ve_{\al}\ve_{\ga}}+
cycle(\al, \beta, \ga)=\nn\\&&=-(i\hbar)^{-3}(\Om_{\al},
\Om_{\beta},
\Om_{\ga})_{\Om}(-1)^{\ve_{\al}\ve_{\ga}}-{2\over3}
(i\hbar)^{-1}[\Om_{\al\beta\ga},\Om].
\e{18}
We may then proceed and derive integrability
conditions for $\Om_{\al\beta\ga}$ which introduces
$\Om$'s with still more indices. Now $\Om$'s
with higher indices are nonzero only
 for theories of rank two and higher which demonstrates
that the starting scenario in section 3 then is incorrect.
$\tilde{\theta}_{\al}$ do not close in this case.\\ \\
{\bf The quantum master equation}\\ \\
The original simple integrability conditions \r{9}
for the operator connections $Y_{\al}$ are
through the basic assumption \r{10} replaced by an
infinite set of integrability conditions for an
infinite set of operators. One may question what we
have gained. However, at this point a
miracle happened. Batalin suggested that there must be
a master equation for all these
integrability conditions and indeed this is the case. This master equation is
\be
&&(S, S)_{\Delta}=i\hbar[\Delta, S],
\e{19}
where
\be
&&\Delta\equiv\Om+\eta^{\al} \pi_{\al}(-1)^{\ve_{\al}}, \quad  \Delta^2=0,
\e{20}
is an extended BRST charge operator ($\pi_{\al}$ are conjugate momenta to
$\phi^{\al}$), and where
\be
&&S(\phi,
\eta)\equiv G+\eta^{\al}\Om_{\al}(\phi)+
\half\eta^{\beta}\eta^{\al}\Om_{{\al}{\beta}}(\phi)(-1)^{\ve_{\beta}}+\nn\\
&&{1\over3!}\eta^{\ga}\eta^{\beta}\eta^{\al}\Om_{{\al}{\beta}{\ga}}
(\phi)(-1)^{\ve_{\beta}+\ve_{\al}\ve_{\ga}}+\nn\\
&&\ldots+
{1\over n!}\eta^{{\al}_n}\cdots\eta^{{\al}_1}\Om_{{\al}_1\cdots
{\al}_n}(\phi)(-1)^{\ve_n}+\ldots
\e{21}
is a master charge operator. $\eta^{\al}$ are new ghosts,
which may be interpreted as ghosts or
superpartners to the group parameters $\phi^{\al}$.
The equivalence between the master equation
\r{19} and the set of integrability conditions for the
$\Om$'s has   been checked up to
third order in
$\eta^{\al}$. In fact, $\Om_{\al\beta\ga}$ in \r{18} and
\r{21} are not the same. They are related
as follows
\be
&&
\Om'_{\al\beta\ga}\equiv\Om_{\al\beta\ga}-{1\over8}
\left\{(i\hbar)^{-1}[\Om_{\al\beta},
\Om_{\ga}](-1)^{\ve_{\al}\ve_{\ga}}+cycle(\al,\beta,\ga)\right\},
\e{22}
where $\Om'_{\al\beta\ga}$ is equal to $\Om_{\al\beta\ga}$ in \r{18}.

The quantum master equation \r{19} is quite different
from the quantum master equation in the
BV-quantization of general gauge theories which is
\be
&&\half(S,S)=i\hbar\Delta
S,
\e{23}
where $S$ is the master action which is a functional of
fields and antifields, and where the
$\Delta$-operator is  the following second order differential operator
\be
&&\Delta\equiv(-1)^{\ve_A}\frac{\stackrel{\ra}
{\partial}}{\partial\phi^A}
\frac{\stackrel{\ra}{\partial}}{\partial\phi^*_{A}}.
\e{24}
The bracket in \r{23} is the classical antibracket.
In contrast the master equation \r{19} is
entirely expressed in terms of operators. We do not
know how the quantum master equations \r{19}
and \r{23} are related if they are related at all.
However, there is a formal similarity between the
$\Delta$-operators \r{20} and \r{24} for $\Om=0$ if one
introduces conjugate momenta to the new
ghosts
$\eta^{\al}$.\\ \\
{\bf Interpretation of the quantum master equation}\\ \\
The new quantum master equation \r{19} encodes generalized Maurer-Cartan
equations.
In the case of Lie groups we have
\be
&&S(\phi,
\eta)\equiv G+\eta^{\al}\la_{\al}^{\beta}(\phi)\pet_{\beta},
\e{25}
which when inserted into \r{19} yields the standard Maurer-Cartan equations
\be
&&\dif_{\al}{\la}_{\beta}^{\ga}-
\dif_{\beta}{\la}_{\al}^{\ga}(-1)^{\ve_{\al}\ve_{\beta}}={\la}^{\eta}_{\al}
{\la}^{\del}_{\beta}
{U}^{\ga}_{{\del}{\eta}}(-1)^{\ve_{\beta}\ve_{\eta}
+\ve_{\ga}+\ve_{\del}+\ve_{\eta}}.
\e{26}
For quasi groups which satisfy the properties \cite{IB}
\be
&&[U^{\ga}_{\al\beta}, U^f_{\del\epsilon}]=0, \quad [[\theta_{\del},
U^{\ga}_{\al\beta}], U^{\eta}_{\epsilon\zeta}]=0,
\e{27}
we still have \r{25} and \r{26}. However, $\la^{\al}_{\beta}$
in \r{25} is then an operator and
$\la^{\al}_{\beta}$ in \r{26} is replaced by a transformed operator (see
\cite{BM1}).\\ \\
{\bf Solution of the quantum master equation}\\ \\
Define transformed master charges and $\Delta$-operators by
\be
&&S(\al)\equiv e^{{i\over\hbar}\al F}
Se^{-{i\over\hbar}\al F}, \quad
\Delta(\al)\equiv e^{{i\over\hbar}\al F}
{\Delta} e^{-{i\over\hbar}\al F},
\e{28}
where $\al$ is a parameter and $F$ an operator.
These transformed operators satisfy then the quantum
master equation
\be
&&(S(\al), S(\al))_{\Delta(\al)}=
i\hbar[\Delta(\al), S(\al)].
\e{29}
If we restrict the transformations such that
 $\Delta(\al)=\Delta$ then $S(\al)$ represents a different
solution to the quantum master
equation than $S$. Now    $\Delta(\al)=\Delta$ requires
$[\Delta, F]=0$ whose solution is
\be
&&F(\phi, \eta)=F(0,0)+(i\hbar)^{-1}[\Delta, \Psi(\phi, \eta)].
\e{30}
For $F(0,0)=0$ we have therefore the following natural invariance
transformation
\be
&&S\;\ra\;S'\equiv \exp{\biggl\{-(i\hbar)^{-2}[\Delta,
\Psi]\biggr\}}\,S\,\exp{\biggl\{(i\hbar)^{-2}[\Delta,
\Psi]\biggr\}}.
\e{31}
Now since $S=G$ is a particular solution of the master equation \r{19}
we conclude that  the general
solution must be
\be
&&S=\exp{\biggl\{-(i\hbar)^{-2}[\Delta,
\Psi]\biggr\}}\,G\,\exp{\biggl\{(i\hbar)^{-2}[\Delta,
\Psi]\biggr\}}.
\e{32}
Notice that this  $S$ satisfies the boundary condition $S(0,0)=G$ as
required by the general form
\r{21}.\\ \\
{\bf Explicit representation of open group transformations}\\ \\
The general solution \r{32} suggests that
\be
&&U(\phi, \eta)\equiv \exp{\biggl\{-(i\hbar)^{-2}[\Delta,
\Psi(\phi, \eta)]\biggr\}}
\e{33}
should represent appropriate group elements for arbitrary open groups.
Group transformed states and operators are then defined by 
($U(0,0)=\bett$)
\be
&&\vb\tilde{A}(\phi, \eta)|\equiv\vb{A}|U^{-1}(\phi, \eta),\quad\tilde{A}(\phi,
\eta)\equiv U(\phi,
\eta){A}U^{-1}(\phi,
\eta).
\e{34}
Notice that the master charge itself is a group transformed ghost charge, which
implies
\be
&&
G|A\hb_g=i\hbar g|A\hb_g\quad\Rightarrow\quad S|\tilde{A}\hb_g=i\hbar
g|\tilde{A}\hb_g,\nn\\
&&[G, A_g]=i\hbar g A_g\quad\Rightarrow\quad
[S, \tilde{A}_g]=i\hbar
g\tilde{A}_g.
\e{35}
The group transformed states and operators \r{34}
satisfy the extended Lie equations
\be
&&\vb
\tilde{A}(\phi,\eta)|\stackrel{\lea}{\tilde{D}}_{\al}\equiv\vb
\tilde{A}(\phi,\eta)|\left(\stackrel{\lea}{\dif_{\al}}-(i\hbar)^{-1}
\tilde{Y}_{\al}(\phi,\eta)\right)=0, \nn\\
&&\tilde{A}(\phi,\eta)\stackrel{\lea}{\tilde{\nabla}}_{\al}\equiv
\tilde{A}(\phi,\eta)\stackrel{\lea}{\dif_{\al}}
-(i\hbar)^{-1}[\tilde{A}(\phi,\eta),
\tilde{Y}_{\al}(\phi,\eta)]=0,
\e{36}
where
\be
&&\tilde{Y}_{\al}(\phi,\eta)= (i\hbar)^{-1}
[\Delta, \tilde{\Om}_{\al}(\phi,\eta)]=   i\hbar U(\phi, \eta)\left(U^{-1}(\phi,
\eta)\stackrel{\lea}{\dif_{\al}}\right).
\e{37}

A natural explicit form of \r{33} is obtained for
the choice $\Psi(\phi, \eta)=\phi^{\al}\pet_{\al}$
which yields
\be
&&U(\phi,
\eta)=\exp{\biggl\{{i\over\hbar}\bigr(\phi^{\al}
\tilde{\theta}_{\al}-\eta^{\al}\pet_{\al}\bigl)\biggl\}}.
\e{38}
Notice that $\tilde{\theta}_{\al}$ and
$\pet_{\al}$ satisfy a closed algebra. For theories with
rank zero and one we may set
$\eta^{\al}=0$ in which case we obtain the group element we started from in section 3.
\newpage
\noindent
{\bf An interpretation}\\ \\
In a BRST quantization of general gauge theories we notice the properties
\be
&&\Om|A\hb=0\quad\Rightarrow\quad\Delta|
\tilde{A}(\phi,\eta)\hb=0,\nn\\
&&[\Om, A]=0\quad\Rightarrow\quad[\Delta,
\tilde{A}(\phi,\eta)]=0.
\e{39}
Since $|
\tilde{A}(0,0)\hb=|A\hb$, $\tilde{A}(0,0)=A$, this may be
interpreted as if we have a BRST theory
with
$\Delta$ as the BRST charge. The extended BRST charge
$\Delta$ looks like a BFV-BRST charge in the
nonminimal sector where
$\phi^{\al}$ are
 Lagrange multipliers and $\eta^{\al}$ antighost momenta, \ie $\Delta$ is the
appropriate BRST charge for path integrals in the Hamiltonian BFV
formulation.\\ \\
{\bf The Sp(2) formulation \cite{BM4}}\\ \\
Arbitrary involutions
\be
&&\{\theta_{\al}(z),
\theta_{\beta}(z)\}=U_{\al\beta}^{\;\;\ga}(z)\theta_{\ga}(z)
\e{40}
may also be embedded into a BRST and antiBRST charge.
In the Sp(2)-version \cite{BLT} these
charges are denoted, $\Om^a$, $a=1,2$, and satisfy at the quantum level
\be
&&[\Om^a, \Om^b]=\Om^{\{a}\Om^{b\}}=0.
\e{41}
Open group transformations may also be derived within this framework.
In fact, all previous results
may be extended to this case \cite{BM4}.
First we imagine that  finite open group transformations are
represented by group elements of the
form
\be
&&U(\phi)=\exp{\{-(i\hbar)^{-2}\ve_{ab}[\Om^b, [\Om^a, R(\phi)]]\}},
\e{42}
where $\ve_{ab}$ is the Sp(2) metric ($\ve_{21}=-\ve_{12}=1$).
Then we assume that group transformed states and operators satisfy
the Lie equations \r{8}.
Due to \r{42} the operator connections $Y_{\al}(\phi)$
should therefore have the form
\be
&&Y_{\al}=(i\hbar)^{-2}\half\ve_{ab}[\Om^b, [\Om^a, X_{\al}(\phi)]].
\e{43}
In order for $Y_{\al}$ to satisfy the same boundary
condition as before, \ie \r{10}, the
Sp(2) charge $\Om^a$ must be given in the non-minimal
sector with dynamical Lagrange multipliers.
Explicitly this means \cite{BLT}
\be
&&\Om^a=\ca^{\al a}\theta_\al+\half\ca^{\beta b}\ca^{\al a} {U}_{\al
\beta}^{\;\;\;\ga}\pet_{\ga
b}(-1)^{\ve_\beta+\ve_\ga}+\ve^{ab}\pet_{\beta b}\la^\beta+\half
\la^\beta\ca^{\al a} U^{\;\;\;\ga}_{\al\beta}\zeta_\ga+\cdots,\nn\\
\e{44}
where $\ca^{\al a}$  and $\pet_{\al
a}$ are Sp(2) ghosts and their conjugate momenta, and where
$\la^{\al}$ and $\zeta_{\al}$ are
the Lagrange multipliers and their conjugate momenta. Their
commutation relations are
\be
&&[\ca^{\al a}, \pet_{\beta
b}]=i\hbar\del^{\al}_{\beta}\del^a_b, \quad [\la^{\al},
\zeta_{\beta}]=i\hbar\del^{\al}_{\beta}.
\e{45}
The expression \r{44} represents the first terms in a
$\ca\pet$- and $\la\zeta$-ordered Sp(2)
 charge. The remaining terms are determined by the conditions \r{41}.
In terms of these
charges the connection operators $Y_{\al}$ are realized
according to the formula \r{43} with the
boundary condition
$X_{\al}(0)=-\zeta_{\al}$.
The integrability conditions \r{9} for $Y_{\al}$ imply here
\be
&&X_{\al}\stackrel{\lea}{\dif_{\beta}}-X_{\beta}
\stackrel{\lea}{\dif_{\al}}(-1)^{\ve_{\al}\ve_{\beta}}+
(i\hbar)^{-3}\half\{X_{\al},
X_{\beta}\}_{\Om}=(i\hbar)^{-1}[X_{\al\beta a}, \Om^a],
\e{46}
where
\be
&&\{f, g\}_{Q}\equiv \ve_{ab} \left[[f, Q^a],
[Q^b, g]\right]
\e{47}
 is a generalized commutator. From \r{46} one may derive
integrability conditions for $X_{\al\beta
a}$ which introduces $X$'s with higher indices etc. However,
also in this case a miracle happens.\\ \\
{\bf The Sp(2) quantum master equation}\\ \\
The integrability conditions for $X_{\al}$, $X_{\al\beta a}$ etc
may be embedded into
the quantum  master equation
\be
&&(S, S)^a_{\Delta}=i\hbar[\Delta^a, S],
\e{48}
where
\be
&&(f, g)^a_{\Delta}\equiv\half \left([f, [\Delta^a, g]]-[g, [\Delta^a,
f]](-1)^{(\ve_f+1)(\ve_g+1)}\right)
\e{49}
is the Sp(2) quantum antibracket which generalizes the
classical Sp(2) antibracket \cite{BLT2}, and
where
\be
&&\Delta^a\equiv\Om^a+\eta^{\al a}
\pi_\al(-1)^{\ve_\al}+\rho^{\al}\xi_{\al
b}\ve^{ab}(-1)^{\ve_{\al}}
\e{50}
is an  extended Sp(2) charge which satisfies
\be
&&[\Delta^a, \Delta^b]=\Delta^{\{a}\Delta^{b\}}=0.
\e{51}
The master charge $S$ in \r{48} is here given by
\be
&&S(\phi,\rho,\eta)=G+\eta^{\al a}\Om_{\al
a}(\phi)+\rho^{\al}\Om_{\al}(\phi)+
\half\eta^{\beta b}\eta^{\al
a}(-1)^{\ve_{\beta}}\Om_{\al\beta
ab}(\phi)+\nn\\&&+\half\rho^{\beta}\rho^{\al}
\Om_{\al\beta}(\phi)+
\rho^{\beta}\eta^{\al
a}\Om_{\al\beta a}(\phi)+\ldots,
\e{52}
where $\phi^{\al}, \rho^{\al}$, and $\eta^{\al a}$ constitute a supersymmetric
set of variables. Their Grassmann parities are
$\ve(\phi^{\al})=\ve(\rho^{\al})=\ve_{\al}$ and
$\ve(\eta^{\al a})=\ve_{\al}+1$.
$\xi_{\al b}$ in \r{50} are conjugate momenta to $\eta^{\al a}$.

The master charge \r{52} is given by a general power
expansion in $\rho^{\al}$ and $\eta^{\al a}$.
The coefficient operators have the following
identifications with the $X$'s in the integrability
conditions
\be
&&X_{\al}=\half\Om_{\al}(-1)^{\ve_{\al}},\nn\\&& X_{\al\beta a}\equiv{1\over6}
\biggl(\Om_{\al\beta a}(-1)^{\ve_{\al}}-\Om_{\beta\al
a}(-1)^{\ve_{\beta}(\ve_{\al}+1)}+(i\hbar)^{-2}\ve_{ab}[[X_{\al},
X_{\beta}], \Om^b]\biggr).\nn\\
\e{53}
The master equation yields furthermore
\be
&&\Om_{\al
a}=\half(i\hbar)^{-1}\ve_{ab}[\Om_{\al}, \Om^b]
\e{54}
and determines $\Om_{\al\beta
ab}$ in terms of $\Om^a, \Om_{\al a}$ and $\Om_{\al\beta a}$.

Even  here there is a similar looking quantum master
equation to \r{48} within the Sp(2) extended
BV quantization \cite{BLT2}. However, it is given
for the master action $S$ and in terms of
classical antibrackets. $\Delta^a$ are second order
differential operators satisfying \r{51} with
only a formal similarity to \r{50} with $\Om^a=0$.\\ \\
{\bf Solution of the Sp(2) quantum master equation}\\ \\
Define as before transformed master charges and
$\Delta$ operators according to
\be
&&S(\al)\equiv e^{{i\over\hbar}\al F}
Se^{-{i\over\hbar}\al F}, \quad
\Delta^a(\al)\equiv e^{{i\over\hbar}\al F}
{\Delta}^a e^{-{i\over\hbar}\al F}.
\e{55}
They obviously satisfy the master equation
\be
&&(S(\al), S(\al))^a_{\Delta(\al)}=
i\hbar[\Delta^a(\al), S(\al)].
\e{56}
In order to derive general solutions of \r{48} we impose the restriction
$\Delta^a(\al)=\Delta^a$ which requires $[\Delta^a, F]=0$. The solution is
\be
&&F(\phi, \eta, \rho)=F(0,0,0)+
\half
\ve_{ab}(i\hbar)^{-2}[\Delta^b, [\Delta^a,
\Phi(\phi,
\eta, \rho)]].
\e{57}
For $F(0,0,0)=0$ we have the natural invariance transformation
\be
&&S\;\ra\;S'\equiv \exp{\biggl\{
-(i\hbar)^{-3}\half\ve_{ab}[\Delta^b,[\Delta^a,
\Phi]]\biggr\}}\,S\,\exp{\biggl\{(i\hbar)^{-3}
\half\ve_{ab}[\Delta^b,[\Delta^a,\Phi]]\biggr\}}.\nn\\
\e{58}
Since $S=G$ is a particular solution of the master equation,
we have therefore the general
solution
\be
&&S= \exp{\biggl\{
-(i\hbar)^{-3}\half\ve_{ab}[\Delta^b,[\Delta^a,
\Phi]]\biggr\}}\, G\,\exp{\biggl\{(i\hbar)^{-3}
\half\ve_{ab}[\Delta^b,[\Delta^a,\Phi]]\biggr\}}.\nn\\
\e{59}\\ \\
{\bf Explicit representation of open group transformations
within the Sp(2) scheme}\\ \\
The result \r{59} suggests that group elements within the
Sp(2) scheme should be defined as
follows ($U(0,0,0)=\bett$)
\be
&&U(\phi, \eta, \rho)\equiv \exp{\biggl\{-\half\ve_{ab}(i\hbar)^{-3}[\Delta^b,
[\Delta^a,
\Phi(\phi,
\eta, \rho)]]\biggr\}}.
\e{60}
Transformed states and operators are then given by
\be
&&\vb\tilde{A}(\phi, \eta, \rho)|\equiv\vb{A}|U^{-1}(\phi, \eta, \rho),\quad
\tilde{A}(\phi,
\eta, \rho)\equiv U(\phi,
\eta, \rho){A}U^{-1}(\phi,
\eta, \rho).\nn\\
\e{61}
It follows then that
the master charge $S$ is   a group transformed ghost charge. The properties
\r{35}
are therefore valid also here.

A natural explicit form of $U(\phi, \eta, \rho)$
is obtained from \r{60} with the choice
$\Phi(\phi, \eta, \rho) =\phi^{\al}\zeta_{\al}$. It is
\be
&&U(\phi,
\eta)=\exp{\biggl\{{i\over\hbar}\bigr(\phi^{\al}\tilde{\theta}_{\al}+\eta^{\al
a}\tilde{\pet}_{\al a}-\rho^{\al}\zeta_{\al}\bigl)\biggl\}},
\e{62}
where
\be
&&\tilde{\theta}_{\al}\equiv(i\hbar)^{-2}\half \ve_{ab}[\Om^b, [\Om^a,
\zeta_{\al}], \quad
\tilde{\pet}_{\al a}\equiv (i\hbar)^{-1}\ve_{ab}[\Om^b,
\zeta_{\al}](-1)^{\ve_{\al}}.
\e{63}
$\zeta_{\al}$, $\tilde{\theta}_{\al}$ and $\tilde{\pet}_{\al a}$ should
satisfy a
closed algebra.\\ \\
{\bf Conclusions}\\ \\
I hope I have been able to convey the message that
it seems possible to develop a group theory even 
for open groups. I expect the results obtained so
far only to constitute a beginning of such a
development. Many consequences remain to be drawn.
There are probably also several reinterpretations
of known properties awaiting to be discovered.


\begin{thebibliography}{Simple}


\bibitem{BM1}I. A. Batalin and
R. Marnelius,       {\it  Phys.
Lett.} {\bf B441},  (1998) 243.

\bibitem{BM2}I. A. Batalin and
R. Marnelius,   {\it Mod. Phys. Lett.} {\bf A14}, (1999) 1643.

\bibitem{BM3}I. A. Batalin and
R. Marnelius,   {\it Nucl. Phys.} {\bf B551},  (1999) 450.

\bibitem{BM4}I. A. Batalin and
R. Marnelius,   {\it Int. J. Mod.
Phys.}  {\bf A15} (2000) 2077.


\bibitem{BFV}I. A. Batalin and G. A Vilkovisky,
 {\it  Phys. Lett.}
 {\bf B69}, (1977) 309;\\
 E. S. Fradkin T. E. Fradkina,  {\it  Phys. Lett.}
 {\bf B72}, (1978) 343;\\
I. A. Batalin and E. S. Fradkin,  {\it  Phys. Lett.}
 {\bf B122}, (1983) 157.

\bibitem{IB}I. A. Batalin,   {\it  J. Math. Phys.}  {\bf 22},  (1981) 1837.


\bibitem{BF}I. A. Batalin and E. S. Fradkin,
 {\it Phys. Lett.} {\bf B128},
(1983) 303;\\ {\it Riv. Nuovo Cim.} {\bf 9},
(1986) 1; {\it Ann. Inst. Henri Poincar\'{e}}
{\bf 49}, (1988) 145.

\bibitem{RM}R. Marnelius,   {\it   Phys. Lett.}  {\bf B99},  (1981) 467.


\bibitem{BM5}
I. A. Batalin and
R. Marnelius,
  {\it  Phys. Lett.} {\bf
            B434}, (1998)  312.

\bibitem{BM6}
I. A. Batalin and
R. Marnelius,
  {\it Theor. Math.  Phys.} {\bf
            120}, (1999)  1115.


\bibitem{BLT}I.A.~Batalin, P.M.~Lavrov, and I.V.~Tyutin,
\JMP\ {\bf 31}, (1990)  6,  2708.

\bibitem{BLT2}I.A.~Batalin, P.M.~Lavrov, and I.V.~Tyutin,
\JMP\ {\bf 31},  1487 (1990); {\bf 32}, (1990)  532,  2513.

\end{thebibliography}
\end{document}